\documentclass[aps,preprint]{revtex4}
\usepackage{amsmath,amssymb,bm}

\renewcommand\d{\partial}
\newcommand\<{\langle}
\renewcommand\>{\rangle}

\newcommand\B{\mathbf{B}}
\newcommand\p{\mathbf{p}}
\newcommand\q{\mathbf{q}}
\newcommand\x{\mathbf{x}}
\newcommand\y{\mathbf{y}}
\renewcommand\j{\mathbf{j}}
\newcommand{\abs}[1]{|#1|}
\newcommand\stru{\rule[-6mm]{0mm}{6mm}}

\begin{document}

\preprint{INT-PUB 05-22}
\title{Response of strongly-interacting matter to magnetic field:\\ some
  exact results}

\author{G.~M.~Newman and D.~T.~Son} 

\affiliation{Institute for Nuclear Theory, University of Washington,
Seattle, Washington 98195-1550, USA}

\begin{abstract}

We derive some exact results concerning the response of
strongly-interacting matter to external magnetic fields.
Our results come from consideration of triangle anomalies in medium.
First, we define an ``axial magnetic susceptibility,'' then we examine 
its behavior in two flavor QCD via response theory. In the chirally restored
phase, this quantity is proportional to the fermion chemical
potential, while in the phase of broken chiral 
symmetry it can be related, through triangle anomalies, to 
an in-medium amplitude for $\pi^0\to2\gamma$.
We confirm the latter result by calculation in a linear sigma model,
where this amplitude is already known in the literature.
\end{abstract}

\maketitle

\section{Introduction}

Much attention has been focused recently on properties of matter at
very high temperatures or baryon density~\cite{Schaefer:2005ff}.  The
interest is driven by the physics of heavy-ion collisions and of the
core of neutron stars.  Most of the discussion is focused on
properties of the system at finite temperature $T$ and chemical
potential $\mu$.  In this paper, we are interested in the properties
of of hot and dense matter under external magnetic field $B$.  This
question is of potential interest for the physics of compact objects.

In contrast to the temperature and chemical potential, the magnetic
field that can be achieved in nature seems to be always small compared
to the strong scale (perhaps as large as $\sim 10^{18}$ G in
magnetars~\cite{Ghosh:2001jv,Malheiro:2004sb}).  This means that the
effect of magnetic field on the medium, in most cases, can be treated
as a small perturbation, and the linear response theory is
appropriate.  In this paper we derive two results related to the
response of a strongly-interacting medium to the external magnetic
field.  We are interested in the axial current created by a uniform
magnetic field $\B$.  For small magnetic fields the axial current is
linear in $\B$, with a proportionality coefficient which we call the
axial magnetic susceptibility $\chi$.  We show that if chiral symmetry
is unbroken, then the value of $\chi$ is equal to the baryon chemical
potential, with a known numerical coefficient.  This result is
universal and receives no correction due to strong coupling.  We also
find that when chiral symmetry is unbroken, this universal result no
longer holds.  However in this case, we derive a relation between the
axial magnetic susceptibility and the in-medium $\pi^0\to2\gamma$
amplitude in spacelike domain (a more precise definition is given
below).  In both cases the results come from the consideration of
triangle anomalies, and hence are reminiscent to those of
Refs.~\cite{Son:2004tq,Metlitski:2005qz}.  The difference between this
work and Refs.~\cite{Son:2004tq,Metlitski:2005qz} is that here we are
interested in the properties of the ambient matter, not of topological
defects as in Ref.~\cite{Son:2004tq,Metlitski:2005qz}.

The paper is organized as followed.  Section~\ref{sec:defs} we define
the quantities of interest in a simplified version of the real world.
In Sec.~\ref{sec:relations} we derive the exact relations.  We
explicitly check this relation in Sec.~\ref{sec:model} in the linear
sigma model.  We extend the results to the real world in
Sec.~\ref{sec:isospin}, and conclude with Sec.~\ref{sec:conclusion}.

\section{Basic definitions}
\label{sec:defs}

For simplicity, we consider QCD with massless $u$ and $d$ quarks and
neglect other quarks.  We shall first assume nonzero baryon chemical
potential, but zero isospin chemical potential.  Moreover, let us
assume that electromagnetism couples to the third component of the
isospin current $\frac12 \bar q \gamma^\mu \tau^3 q$, but not a linear
combination of isospin and baryon current as in the real world.  We
shall modify our result to the real world later.  We will assume no
superconductivity, so the magnetic field can penetrate the matter
without being confined in magnetic flux tubes (this may require that
we work at sufficiently high temperatures).

Let us first define the axial magnetic susceptibility.  Assume we have
a medium at temperature $T$ and baryon chemical potential $\mu$ in an
uniform magnetic field $\B$.  The magnetic field induces an axial
current in the medium,
\begin{equation}\label{chi}
  \< \j^5 \> = \chi(T, \mu) \B
\end{equation}
where, in terms of quark fields,
\begin{equation}
  j^{5\mu} = \frac12(\bar u \gamma^\mu\gamma^5 u 
  - \bar d\gamma^\mu\gamma^5 d ).
  \label{ac-def}
\end{equation}

That $\chi$ is nonzero is permitted by symmetries.  With respect to
parity, both $\j^5$ and $B$ are axial vectors.  Note that $B$ and
$A_i$ has different $C$ parity, so $\chi$ must be an odd function of
the baryon chemical potential $\mu$.  In particular, $\chi=0$ at zero
chemical potential.  In the model described in Sec.~\ref{sec:model},
$\chi$ is proportional to $\mu$ at small $\mu$.

If the medium consists of non-relativistic nucleons, then the axial
current is proportional to the nucleon spin.  The coefficient $\chi$
therefore has a simple physical interpretation as the spin
polarizability of the medium (more precisely, the the difference
between proton and neutron spin polarizability).  When nucleons are
relativistic, it becomes impossible to separate nucleon spin from
nucleon total angular momentum.  However even in this case, the axial
magnetic susceptiblity is still a well-defined concept.  In fact, this
axial current interacts, through $Z^0$ exchange, with neutrinos and
modify their dispersion relation.

We also define the in-medium coupling of the neutral pion to two photons,
$g_{\pi^0\gamma\gamma}$, in the chirally broken phase.  In vacuum, the
anomalous coupling of a pion to external gauge fields is given by a
term in the chiral Lagrangian,
\begin{equation}
  -\frac1{8\pi^2} \epsilon^{\mu\nu\alpha\beta}\d_\mu \phi
  A^B_\nu F_{\alpha\beta}, \qquad \phi = \frac{\pi^0}{f_\pi},
\end{equation}
where $A^B_\mu$ is the gauge potential coupled to the baryon current
(note that $A_\mu$ couples to the isospin current).  The dimensionless
field $\phi$ is normalized to have periodicity $2\pi$.
At finite temperature, the coupling is more subtle.  As
noted in Ref.~\cite{Gelis}, there is an ambiguity with the zero
momentum limit.  We choose the following definition.  Let us look at
the free energy of a \emph{static} field configuration where the
$\pi^0$ field changes slowly in space, in the presence of a background
static magnetic field $\B$ and static baryon scalar potential $A_0^B$.
The free energy has the form
\begin{equation}\label{F}
  F = \frac{f_s^2}2 (\d_i \phi)^2 - g_{\pi^0\gamma\gamma} 
      \frac1{8\pi^2f_s}\epsilon^{ijk} \d_i\phi A^B_0 F_{jk}.
\end{equation}
Here $f_s$ is the spatial pion decay constant, and
$g_{\pi^0\gamma\gamma}$ will be called the $\pi^0\to2\gamma$
amplitude.  The free energy~(\ref{F}) can be thought of as arising
from integrating out all degrees of freedom of QCD except the
Goldstone boson, and restricting to the lowest Matsubara frequency
$\omega=0$.  At zero temperture $f_s=f_\pi$ and
$g_{\pi^0\gamma\gamma}=1$, but in general both $f_s$ and
$g_{\pi^0\gamma\gamma}$ are functios of temperature and baryon
chemical potential.  We also know that $f_s\to0$ at the second-order
chiral phase transition, with the critical exponent of Josephson's
scaling~\cite{Son:2001ff}.  The way we define $g_{\pi^0\gamma\gamma}$
corresponds to the $\pi^0\to2\gamma$ amplitude in the spacelike region
of Ref.~\cite{Gelis}.

To summarize, our results are
\begin{itemize}
\item When the point $(T,\mu)$ lies in the chirally restored phase,
$\chi$ is directly proportional to the chemical potential,
\begin{equation}
  \chi = \frac1{4\pi^2}\mu. \label{result-1}
\end{equation} 
The numerical coefficient $1/(4\pi^2)$ is exact and is related to
triangle anomaly.  

\item When chiral symmetry is sponteneous broken, the relation between
$\chi$ and the anomaly is lost.  However, there is an exact equation
relating the susceptibility and the in-medium amplitude of
$\pi^0\to2\gamma$.  Namely,
\begin{equation}
  4\pi^2\frac{d\chi}{d\mu} 
  + g_{\pi^0\gamma\gamma}(T,\mu) = 1. \label{result-2}
\end{equation} 
Here $g_{\pi^0\gamma\gamma}(T)$ is the $\pi^0\to2\gamma$ amplitude
defined above.
\end{itemize}

We note that the first part of our results, which concerns the
chirally restored phase, has been observed in Ref.~\cite{Metlitski:2005pr}.
In addition, it is only a slight variation on the result that is
found in Ref.~\cite{Alekseev:1998ds}, where it was determined that
the magnetic susceptibility of the electric current is proportional to
a chemical potential for fermion chirality. This relation can also be 
checked explicitly in models of strongly-interacting field theory with 
gravity dual description~\cite{anom-hydro}.  However, to our knowledge,
the case with spontaneous breaking of chiral symmetry has never been 
considered before; hence the second part of our results is new. 
We first show the validity of these relations in a general setting. 
Then we shall verify them explicitly in a model with anomaly, namely 
the linear sigma model.

\section{Exact relations}
\label{sec:relations}

To derive the exact relations, we consider a three-point correlation
function of the axial current $j^5_\mu$, the isospin current $j^\mu$,
and the baryon current $B^\mu$,
\begin{equation}
  i \Gamma^{\mu\nu\lambda}(p,q) = \int\!d^4x\, d^4y\, 
  e^{ip\cdot x+iq\cdot y} \< j^{5\mu}(x) j^\nu(y) B^\lambda(0)\>,
\end{equation}
where $j^{5\mu}$ is defined in Eq.~(\ref{ac-def}) and other currents are
defined as follows
\begin{equation}
  j^\mu = \frac12(\bar u \gamma^\mu u - \bar d\gamma^\mu d),\qquad
  B^\mu = \frac13(\bar u \gamma^\mu u + \bar d\gamma^\mu d).
\end{equation}
We can always define the correlator so that the triangle anomaly
resides entirely in the derivative of the axial current:
\begin{align}
  p_\mu \Gamma^{\mu\nu\lambda}(p,q) &= -\frac1{4\pi^2}
  \epsilon^{\nu\lambda\alpha\beta} p_\alpha q_\beta, \label{anomaly}\\
  q_\nu \Gamma^{\mu\nu\lambda}(p,q) &= (p_\lambda+q_\lambda)
   \Gamma^{\mu\nu\lambda}(p,q) = 0.
\end{align}

In this paper we will be interested only in static (time-independent)
problems, therefore we shall set $p_0=q_0=0$.  Moreover, the baryon
chemical potential couples to the zeroth component of $B_\lambda$.
Thus the quantity of interest for us will be
\begin{equation}
  i \Gamma^{ij0} (\p, \q) = \int\!d^4x\, d^4y\,
  e^{-i\p\cdot\x-i\q\cdot\y}
  \< A^i(x) V^j(y) B^0(0)\>.
\end{equation}

On the other hand, the axial magnetic sucsceptibility $\chi$ is
related to the low-momentum behavior of a two-point function.  Indeed,
the axial current created by a background electromagnetic field is
\begin{equation}
  \< j^{5\mu}(x) \> = -i \int\!d^4y\, G_{5I}^{\mu\nu}(x-y) A_\nu(y),
\end{equation}
where
\begin{equation}
  G^{\mu\nu}_{5I}(x-y) = \< j^{5\mu}(x) j^\nu(y)\>.
\end{equation}
In order to reproduce Eq.~(\ref{chi}), the infrared behavior of the
$G_{5I}$ correlator must be as follows:
\begin{equation}\label{2pointchi}
  \int\!d^4x\, e^{-i\p\cdot\x}\< j^{5i}(x)j^k(0) \>_\mu =
  \chi \epsilon^{ijk} p_j  + O(p^2).
\end{equation}

In Eq.~(\ref{2pointchi}) we emphasize that the average is taken at
nonzero chemical potential $\mu$.  Differentiating
Eq.~(\ref{2pointchi}) with respect to $\mu$ we get
\begin{equation}\label{Gammachi}
  \Gamma^{ik0} (\p, -\p) = 
  - \frac{\d\chi(\mu)}{\d\mu} \epsilon^{ijk} p^j + O(p^2).
\end{equation}

Let us now look at the structure of the correlator $\Gamma_{ij0}(\p,
\q)$ in the regime of small $\p$ and $\q$.  It changes sign under
parity.  This static correlator is not singular in the chirally
restored phase, and may have a pion pole in the chirally broken phase.
In light of this, the general form of the three point function is,
\begin{equation}\label{Gamma-2contr}
  \Gamma^{ij0}(\p,\q) = C_1\epsilon^{ijk}q^k 
  + C_2\frac{p^i}{p^2}\epsilon^{jkl} q^k p^l.
\end{equation}
Since $\Gamma$ is even under $C$, the dimensionless constants $C_1$
and $C_2$ must both be even under $C$ parity, with $C_2$ vanishing in
the chirally restored phase.

From the relation of triangle anomaly~(\ref{anomaly}) we find
\begin{equation}\label{sum}
  C_1 + C_2 = \frac1{4\pi^2}.
\end{equation}
Consider first the case when chiral symmetry is restored.  Then as the
singular term in Eq.~(\ref{Gamma-2contr}) is absent and
$C_1=1/(4\pi^2)$.  But by comparing Eq.~(\ref{Gamma-2contr}) with
Eq.~(\ref{Gammachi}), we find
\begin{equation}
  \frac{\d\chi(\mu)}{\d\mu} = C_1 = \frac1{4\pi^2}.
\end{equation}
Requiring $\chi$ to be an odd function of $\mu$, we determine
\begin{equation}
  \chi(\mu, T) = \frac\mu{4\pi^2},
\end{equation}
which is the first part of our result.

Now we turn to the chirally broken phase.  The singular term in
Eq.~(\ref{Gamma-2contr}) comes from the Feynman diagram with an
intermediate pion line, which can be computed using the free
energy~(\ref{F}) as an effective Lagrangian.  This leads us to
\begin{equation}
  C_2 = \frac{g_{\pi^0\gamma\gamma}}{4\pi^2}.
\end{equation}
Equation~(\ref{sum}) then implies
\begin{equation}
  4\pi^2\frac{\d\chi}{\d\mu} + g_{\pi^0\gamma\gamma} = 1,
\end{equation}
which is the second part of our result.

\section{Example: Linear Sigma Model}
\label{sec:model}

Since QCD is strongly coupled for temperatures below the chiral phase
transition, we cannot directly check the formula~(\ref{result-2})
there (although it should be possible to verify it in the high-density
phase of three-flavor QCD, where chiral symmetry is broken at weak
coupling).  We shall instead verify this formula in a weakly coupled
field theory with anomaly, namely the linear sigma model.  This model
was employed before to understand the effects of temperature on
anomaly~\cite{Itoyama}. Furthermore, the finite-temperature
$\pi^0\to2\gamma$ amplitude has been computed in this model in various
kinetimatic limits, including the limit where the outgoing photons are
at zero frequency~\cite{Pisarski,Gelis}.  Thus, we can confirm our
result~(\ref{result-2}) by calculating the axial magnetic
susceptibility in this model.

The model is given by the Lagrangian
\begin{equation}
\mathcal{L} = 
\bar{Q}(i\gamma^{\mu}D_{\mu} - g\sigma 
  + i\bm{\tau}\cdot\bm{\pi}\gamma^5)Q +
\frac{1}{2}(D_{\mu}\sigma D^{\mu}\sigma 
 + D_{\mu}\bm{\pi}\cdot D^{\mu}\bm{\pi}) +
\frac{\mu^2}{2}(\sigma^2+\bm{\pi}^2) 
 - \frac{\lambda}{4}(\sigma^2+\bm{\pi}^2)^2.
\end{equation}
The couplings $g$ and $\lambda$ are small.  The expectation value of
$\sigma$ is $v$, which is temperature dependent (being equal to 
$\sqrt{\mu^2/\lambda}$ in vacuum.) The covariant derivative, 
\(D_\mu = \d_\mu+iqA_\mu\), is that of $U(1)_{\rm EM}$. In the 
phase with chiral symmetry breaking, fermions (``constituent quarks'') 
have mass $m=gv$.  We shall calculate the axial current induced by 
a baryon chemical potential $\mu_B$ on the background of a constant 
magnetic field, using the single-particle Hamiltonian in the regime 
where $T \gg \mu, m, \sqrt{eB}$.

Placing a static, homogeneous magnetic field pointing in the $\hat{z}$
direction can be accomplished by means of the vector potential.
$A^{\mu} = (0,0,Bx,0)$. The fermion spectrum can be found by
solving the Dirac equation \( i \bm{\gamma}\cdot\mathbf{D} Q = E
Q\) with \(D^{\mu}=\partial^{\mu}-iq\cal{A}^{\mu}\), where, $q$ is the
electric charge of the quark $Q=(u,d)$, $q_u=-q_d=\frac12$.  
The axial-vector current can then be directly calculated as a thermal 
expectation value of~(\ref{ac-def}).

Within this setup it is convenient to parametrize the quark wave functions as,
\begin{equation}
  Q(x,y,z) = \left( \begin{array}{c} \psi_{1}(x) \\ \psi_{2}(x) \\ 
  \phi_{1}(x) \\ \phi_{2}(x) \end{array}   \right)e^{i(p_{y}y+p_{z}z)}.
\end{equation}
Because of our choice of coordiante system, ``$p_y$'' will always appear 
in the same component of the Dirac equation as ``$eBx$'' in the linear 
combination $p_y + eBx$. Thus, by making a convenient coordinate shift, 
\(x \rightarrow x - \frac{p_y}{qB}\) we can eliminate the explicit 
appearance of ``$p_y$'' in the Dirac equation. We have four first order 
coupled DEs, from which we can obtain two second order equations for 
$\psi_{2}$ and $\phi_{2}$ separately:
\begin{align}
[\nabla^{2} - (qBx)^{2} + E^{2} - p_{z}^{2} - m^{2} + qB]\psi_{2}(x) &= 0,\\
[\nabla^{2} - (qBx)^{2} + E^{2} - p_{z}^{2} - m^{2} + qB]\phi_{2}(x) &= 0.
\end{align}
The other components can be expressed via $\psi_2$ and $\phi_2$, as
\begin{align}
\psi_{1} &= i[\nabla + qBx]\frac{(E-p_{z})\psi_{2}
   -m\phi_{2}}{E^{2}-p_{z}^{2}-m^{2}},\\
\phi_{1} &= -i[\nabla + qBx]\frac{(E+p_{z})\phi_{2}
   -m\psi_{2}}{E^{2}-p_{z}^{2}-m^{2}}.
\end{align}
The components $\psi_2(x)$ and $\phi_2(x)$ are clearly the
eigenfunctions of a harmonic oscillator,
while the other components are obtained by acting a lowering operator
on linear combinations of $\psi_2$ and $\phi_2$. 
The energy eigenvalues are thus the familiar 
Landau levels, \( E = \pm\sqrt{p_{z}^{2}+m^{2}+2nqB} \). 
The constituent quark wave function, properly normalized, is
\begin{equation}
Q(x,y,z) =  \frac{1}{\sqrt{2^{n+2}n!\sqrt{\pi}}}\left( \begin{array}{c} 
i\frac{\alpha(E+p_{z})-\beta m}{\sqrt{qB}}\mathcal{H}_{n-1}(\sqrt{qB}x) \\ 
\alpha\mathcal{H}_{n}(\sqrt{qB}x) \\
i\frac{\alpha m - \beta(E-p_{z})}{\sqrt{qB}}\mathcal{H}_{n-1}(\sqrt{qB}x) \\
\beta\mathcal{H}_{n}(\sqrt{qB}x)
\end{array} \right)e^{-qBx/2}e^{i(p_{y}y+p_{z}z)}.
\end{equation}
The Landau levels, $n$, replace the $p_x$ quantum number.

The $\mathcal{H}$ are the Hermite polynomials, with the addition that, 
for $n=0$ we define \(\mathcal{H}_{-1}\equiv 0\).  For each $n>0$
there are two eigenstates with each choice of the sign of $E$:
\begin{equation}
  \left( \begin{array}{c} \alpha\\ \beta \end{array}\right) =
  \left(\frac{qB}{16\pi}\right)^{1/4}
  \left( \begin{array}{c} 
     \dfrac{m-\sqrt{2nqB}}{\sqrt{E(E-p_{z})}} \stru\\
     \sqrt{\dfrac{E-p_{z}}{E}} \end{array}
   \right), \quad
   \left(\dfrac{qB}{16\pi}\right)^{1/4}
   \left( \begin{array}{c} \sqrt{\dfrac{E+p_{z}}{E}} \stru\\
         \dfrac{m+\sqrt{2nqB}}{\sqrt{E(E+p_{z})}} \end{array}
   \right).
\end{equation}
Meanwhile, for $n=0$ there is only one positive energy eigenstate and one
negative energy eigenstate,
\begin{equation}
  \left( \begin{array}{c} \alpha \\ \beta \end{array}\right) = 
  \left(\frac{qB}{4\pi}\right)^{1/4}
  \left( \begin{array}{c}
     \sqrt{\dfrac{E+p_{z}}{E}}\stru \\ \sqrt{\dfrac{E-p_{z}}{E}}
  \end{array}\right).
\end{equation}

Now we calculate,
\begin{equation}
\< A^{3}_{z} \> \equiv \<\bar{Q}\gamma_{z}\gamma^{5}Q\>_{\mu,T} 
- \<\bar{Q}\gamma^{z}\gamma^{5}Q\>_{0,T}.
\end{equation}
by filling up the fermion energy levels with the Fermi-Dirac
distribution function,
\begin{equation}
\< \j^5 \> = \frac{q\B}{2\pi}\sum_{n,\lambda}
  \sum_{sgn(E)}\int\frac{dp_{z}}{2\pi}
[\abs{\psi_{1}}^{2}-\abs{\psi_{2}}^{2}+\abs{\phi_{1}}^{2}-\abs{\phi_{2}}^{2}]
[\frac{1}{e^{\beta(E-\mu_q)}+1}-\frac{1}{e^{\beta E}+1}]\hat{z}.
\end{equation}
Here $\mu_q=\frac13\mu$ is the quark chemical potential, and
the prefactor $(qB)/(2\pi)$ arises from the degeneracy of the
energy eigenstates in $p_y$.
The sum is taken over all Landau levels $n$, both
sign of energy $E$ and all polarization $\lambda$.  For regularization
we also subtracted the value of the axial current at $\mu=0$, which is
zero by $C$ parity.  For $n>0$ the sum over polarizations read
\begin{equation}
  \sum_\lambda [\abs{\psi_{1}}^{2}-\abs{\psi_{2}}^{2}+\abs{\phi_{1}}^{2}
-\abs{\phi_{2}}^{2}] =
  \frac{1}{4nqB}\{ 2m^{2} 
  - 4\frac{p_{z}}{E}(m^{2}+2nqB -
  4\frac{m^{2}\sqrt{2nqB}}{m^{2}+2nqB}) - 2m^{2} \},
\end{equation}
which is an odd function of $p_z$ and contributes nothing to the axial
current after integration over $dp_z$.  It is very easy to see that
the $n>0$ Landau levels do not contribute to the axial current in two
limits, when the fermions are massless and when the fermions are
nonrelativistic (very massive), so it it not entirely surprising that
they do not contribute for any value of $m$.

The contribution from the lowest Landau level is considerably simpler,
since here
\begin{equation}
[\abs{\psi_{1}}^{2}-\abs{\psi_{2}}^{2}+\abs{\phi_{1}}^{2}-\abs{\phi_{2}}^{2}] =
\frac{\abs{\alpha}^{2}+\abs{\beta}^{2}}{\sqrt{eB}} = 1.
\end{equation}

All that remain is to calculate the integral over the statistical
factor.  This can be done analytically at large temperatures by 
expanding to first order in the small quantities, 
$\mu/T$ and $m^{2}/T^{2}$. For the sake of brevity
we use 
$n(\xi)\equiv(e^{\xi}+1)^{-1}$, and 
$n'(\xi)=\partial_{\xi}n(\xi)$, with $\xi\equiv{p_{z}}/{T}$.  In
particular,
\begin{equation}
\frac{1}{e^{\beta(E-\mu_q)}+1}-\frac{1}{e^{\beta E}+1} \approx 
  -\beta\mu_q\frac{d}{d(\beta E)}n(\beta E)
  = -\beta\mu_q \left[n'(\xi)+\frac{m^2}{T^2}
\frac{1}{2\xi}n''(\xi)\right]. \label{stat-int}
\end{equation}
The sum over flavors yields
$q_u - q_d = 1$
in place of ``$q$'' in the prefactor $qB/(2\pi)$.
Also, a factor of $N_c=3$ is gained from the sum over colors. 
Thus we find
\begin{equation}
\< \j^5 \> = 
\frac{\mu\B}{4\pi^{2}}\left[1 - \frac{m^2}{T^2}\int^{\infty}_{0}
d\xi\frac{n''(\xi)}{2\xi} + \mathcal{O}\left(\frac{m^4}{T^4}\right) 
+ \mathcal{O}\left(\frac{\mu}{T}\right)\right]. \label{result}
\end{equation}
Performing the integral as in Ref.~\cite{Gelis}, we obtain
\begin{equation}
\frac{\d\chi}{\d\mu} = 
\frac1{4\pi^{2}} \left[1 
  -\frac{7\zeta(3)}{4\pi^{2}} \frac{m^2}{T^2} 
+ \mathcal{O}(\frac{m^4}{T^4}) + \mathcal{O}(\frac{\mu}{T})\right].
\end{equation}
Note that we recover Eq.~(\ref{result-1}) for $m=0$.

On the other hand, the result of Refs.~\cite{Pisarski,Gelis}, in our
languague, corresponds to 
\begin{equation}
   g_{\pi^0\gamma\gamma} = \frac{7\zeta(3)m^{2}}{4\pi^2 T^{2}},
\end{equation}
We see that
\begin{equation}
  4\pi^2 \frac{\d\chi}{\d\mu} + g_{\pi^0\gamma\gamma} = 1,
\end{equation}
in accordance with Eq.~(\ref{result-2}).

\section{Isospin chemical potential, real-world EM coupling}
\label{sec:isospin}

In general, the system could contain an isopsin chemical potential as well. 
This section expresses the results of the last two sections, as they 
partain to two flavor QCD with non-zero $\mu_I$ and $\mu_B$, in an applied 
magnetic field coupling properly to $U(1)_{\rm EM}$.
The $U(1)_{\rm EM}$ vector current can be expressed in the flavor basis, 
as a linear combination of the isospin and baryonic generators. 
Specifically, 
\( j^\mu_{\rm Q} \equiv j^\mu_{\rm EM} = e(\frac1{2}j^\mu_{\rm B} + j^\mu_{\rm I}) \).
We now expect,
\begin{equation}
\< \j^5 \> = \chi(T,\mu_{\rm I},\mu_{\rm B})\B_{\rm Q}.
\end{equation}
We will see that $\chi(T,\mu_{\rm I},\mu_{\rm B})$ is a linear combination 
of $\mu_{\rm I}$ and $\mu_{\rm B}$.

In the chirally symmetric phase, we can follow section~\ref{sec:relations}
but replacing the $A^\nu$ coupling to isospin with one coupling 
to electric charge, $A^\nu_{\rm Q}$. Then the response of the chiral 
current can be expressed in terms of the correlator, $G_{5Q}$ as
\begin{equation}
\< j^{5\mu}(x) \> = -i\int\!d^4y\,G^{\mu\nu}_{5Q}(x-y)A_{\nu Q}(y).
\end{equation}
 For this system, we must have
\begin{equation}
\int\!d^4x\,e^{ip\cdot x}G^{ik}_{5Q}(x) = \chi\epsilon^{ijk}p_{j} + O(p^2).
\end{equation}
Now, two different anomaly relations -- one for $\<j^{5i}j^{j}_{Q}B^{0}\>$ 
and one for $\<j^{5i}j^{j}_{Q}V^{0}\>$ -- give respectively,
\begin{equation}
 \frac{\d\chi}{\d\mu_B} = \frac1{4\pi^2}, \quad \frac{\d\chi}{\d\mu_I} = \frac1{8\pi^2}.
\end{equation}

For the phase with broken chiral symmetry, the analysis of section~\ref{sec:relations}
still holds as well, but with the distinction that there would be 
two separate diagrams with propagating pions, necessitating the 
introduction of two amplitudes, $g_{\pi^0\gamma I}$ and $g_{\pi^0\gamma B}$ 
defined analogously with $g_{\pi^0\gamma\gamma}$. Again, restricting the 
anomaly relation to different space time components results in two 
different equations,
\begin{equation}\label{sumrule-i}
 8\pi^2\frac{\d\chi}{\d\mu_I} + g_{\pi^0\gamma I} = 1, 
 \quad 4\pi^2\frac{\d\chi}{\d\mu_B} + g_{\pi^0\gamma B} = 1.
\end{equation}
The $\pi^0\to\gamma\gamma$ amplitude can be defined as a linear
combination of $g_{\pi^0\gamma I}$ and $g_{\pi^0\gamma B}$:
\begin{equation}
  \frac{g_{\pi^0\gamma\gamma}}{4\pi^2} = 
  \frac12 \frac{g_{\pi^0\gamma B}}{4\pi^2} + 
  \frac{g_{\pi^0\gamma I}}{8\pi^2}.
\end{equation}
($g_{\pi^0\gamma\gamma}$ is normalized to 1 at zero temperature.)
From Eqs.~(\ref{sumrule-i}) we find
\begin{equation}\label{pigg}
  4\pi^2\left(\frac{\d\chi}{\d\mu_I} + \frac12 \frac{\d\chi}{\d\mu_B}\right)
  + g_{\pi^0\gamma\gamma} = 1.
\end{equation}

This result could again be confirmed for the sigma model by noting that
the sum over flavors should be performed with \( \mu_u \ne \mu_d \),
resulting in ``$q$'' in the prefactor being replaced by $q_u\mu_u - q_d\mu_d$.
Envoking \(\mu_B = \frac{1}{3}(\mu_u + \mu_d)\) and \(\mu_I = \frac{1}{2}(\mu_u - \mu_d)\) 
we find 
\begin{equation}
q_u\mu_u - q_d\mu_d = \frac{e}{3}\left(\mu_B + \frac12\mu_I\right).
\end{equation}
Thus, 
\begin{equation}
\chi(T,\mu_{\rm I},\mu_{\rm B}) = \frac{e}{4\pi^2}
  \left(\mu_B + \frac12\mu_I\right)\left[1 
  -\frac{7\zeta(3)}{4\pi^{2}} \frac{m^2}{T^2} 
+ \mathcal{O}\left(\frac{m^4}{T^4}\right) 
   + \mathcal{O}\left(\frac{\mu_B}{T}\right) 
  + \mathcal{O}\left(\frac{\mu_I}{T}\right)\right].
\end{equation}
From the results of Refs.~\cite{Pisarski,Gelis}, $g_{\pi^0\gamma\gamma}=
7\zeta(3)m^2/(4\pi^2T^2)$, one can verify Eq.~(\ref{pigg}).

\section{Conclusions}
\label{sec:conclusion}

We have shown that there is a close connection between the response of
strongly interacting matter on external magnetic field and
the axial anomaly. By considering the properties of the three point 
function of isovector, axial isovector, and baryon currents in the presence
of non-vanishing baryon chemical potential and QED magnetic field we were
able to show that the axial-magnetic susceptibility, $\chi(\mu,T)$ is 
directly proportional to the baryon chemical potential
in the absence of Goldstone modes carrying the axial current.
Alekseyev \emph{et. al.} previously found the same to be true for massless 
QED, using quite general methods~\cite{Alekseev:1998ds}.
This result follows from the fact that the anomaly coefficient, 
which determines the constants value, is not renormalized, and 
receives no finite-temperature contibution. In the presence 
of massless pions, this direct relation no longer holds but one still can
relate $\chi(\mu,T)$ to the anomaly coefficient through the 
amplitude for $\pi^0\rightarrow\gamma\gamma$. 
We confirmed the second relation for a particular 
case of weakly-coupled linear sigma model. 

These results may be applicable to the study of compact objects such as 
neutron stars, where both baryonic chemical potential, and magnetic field 
may be large. Specifically, the self energy of neutrinos is affected by 
interaction with the axial isovector current through $Z^0$ 
exchange~\cite{Elmfors:1996gy,Kusenko:1996sr}. Calculation of
$g_{\pi^0\gamma\gamma}(\mu,T)$ in nuclear matter is complicated by 
the need to deal with singularities arising from particle-hole 
interactions, but our results could be easily employed to find this 
neutrino self energy contribution in deep cores of neutron stars, if
chiral symmetry is restored there. In a strong magnetic field the 
momentum distribution of neutrinos is asymmetric already in equilibrium,
so they will stream out in asymmetric fashion, giving rise to a small
contribution to pulsar velocities.  The presence of an axial current 
in matter will not affect oscillations between active neutrinos, but does
change the oscillations between an active neutrino and a sterile one.

We end the paper by noting that currently we lack an understanding of
the critical behavior of $g_{\pi^0\gamma\gamma}$ near the second-order
chiral phase transition.  While it is natural that this coefficient
goes to zero smoothly at the phase transition, the question about
the critical exponent remains open.


This work is supported, in part, by DOE grant DE-FG02-00ER41132.
D.T.S. is supported, in part, by the Alfred P.~Sloan Foundation.  We
thank A.~R~Zhitnitsky for comments on an earlier draft of this
manuscript.

\end{document}